# Mapping the Milky Way: William Herschel's Star-Gages

Todd Timberlake, Berry College, Mount Berry, GA

Abstract:  In 1785 astronomer William Herschel mapped out the shape of the Milky Way star system using measurements he called "star-gages." Herschel's star-gage method is described in detail, with particular attention given to the assumptions on which the method is based. A computer simulation that allows the user to apply the star-gage method to several virtual star systems is described. The simulation can also be used to explore what happens when Herschel's assumptions are violated. This investigation provides a modern interpretation for Herschel's map of the Milky Way and why it failed to accurately represent the size and shape of our galaxy.

William Herschel (Figure 1) is rightfully known as one of the greatest astronomers of all time.  Born in Hanover (in modern Germany) in 1738, Herschel emigrated to England in 1757 and began a successful career as a professional musician.  Later in life Herschel developed a strong interest in astronomy.  He began making his own reflecting telescopes in 1774, and soon his telescopes were recognized as the finest in the world.  It was through one of his homemade telescopes, a Newtonian reflector with a focal length of seven feet and an aperture of 6.2 inches that Herschel first spotted the planet Uranus in 1781.  The discovery of a new planet catapulted Herschel to fame and secured him a position as personal astronomer to King George III.

Although Herschel is most famous for discovering a planet, it was the stars and nebulae to which he devoted most of his astronomical attention.  From 1784 to 1785, ably assisted by his sister Caroline, Herschel attempted to map out the shape of the Milky Way star system.  His cross-section of the Milky Way, shown in Figure 2, is often reproduced in astronomy textbooks and was even featured as one of the 100 greatest scientific discoveries in a recent television show.[1] Most textbooks, however, provide only a brief mention of *how* Herschel mapped out the Milky Way.   Fewer still discuss the problems, recognized by Herschel himself, with his method.[2]

The purpose of this paper is to present a detailed and accurate description of Herschel's approach to mapping the Milky Way, and to present a computer simulation that allows the user to apply Herschel's method to map out various virtual star systems.  The simulation illustrates the soundness of Herschel's method, provided his fundamental assumptions are correct.  More importantly, the simulation shows what happens when those assumptions are violated.  This simulation can be used as part of an activity that will help students not only learn about Herschel's map of the Milky Way, but also about an important aspect of the nature of science.

## Herschel's Star-Gages

Herschel called his measurements of the Milky Way "star-gages." His method relies on two fundamental assumptions:
1. Stars are distributed more or less uniformly within the Milky Way system and are not found beyond the boundaries of that system.
2. The telescope used for the star-gages is capable of resolving all stars within the Milky Way system.

Note that some textbooks claim that Herschel's star-gage method is based on the assumption that all stars have the same luminosity as the Sun. Although Herschel made use of this assumption in other work, it is not needed for his star-gages.

Herschel recognized that the region of space that he could see through his telescope formed a cone with the vertex at the telescope. The telescope's field of view (the angular separation between opposite points within the viewing area of the telescope) is twice the apex angle θ of the cone (see Figure 3). The cone extends outward, encompassing all of the stars that can be seen in the telescope. We can imagine this cone extending to the edge of the Milky Way. There is no need to extend the cone beyond this point, because (according to assumption 1) there are no stars beyond the edge of the Milky Way system.

By assumption 1, the stars within the Milky Way system are distributed uniformly and therefore the number of stars seen through the telescope is proportional to the volume of the cone:

$$V = \left(\frac{1}{3}\right)\pi h^3 \tan^2 \theta, \qquad (1)$$

where $h$ is the distance from the telescope to the edge of the Milky Way. By assumption 2, the telescope is capable of resolving all of the stars in this cone. If we define a *stellar unit* as the average distance between neighboring stars in the Milky Way, then the number of stars $N$ seen in the telescope is equal to the volume $V$ of the cone in cubic stellar units. If we know the telescope's field of view then we know θ and we can use Equation (1) to determine $h$ in stellar units:

$$h = \sqrt[3]{\frac{3N}{\pi \tan^2 \theta}}. \qquad (2)$$

Herschel described his star gage method in 1784.[3] In 1785 he published his results:[4] star-gages of 683 regions in the sky, forming a circle running through the galactic poles and crossing the plane of the Milky Way at right angles near the star Sirius (Alpha Canis Majoris) and again near the star Altair (Alpha Aquilae). The star counts were made with Herschel's "large" 20-foot reflector (a Newtonian reflector with a focal length of 20 feet and an aperture of 19 inches, shown in Figure 4.) In regions where he saw few stars, Herschel averaged the number of stars over ten or more fields. In regions where there were many stars he sometimes only counted stars in one half, or even one fourth, of the field and scaled the result accordingly.

Herschel's star-gages are summarized in the diagram shown in Figure 2. The cross-section of the Milky Way is roughly elliptical in shape, much more extended along the galactic plane (to the left and right in the diagram) than toward the poles (the galactic North pole is toward the top of the diagram). A great cleft in the system appears on the left side, in the direction of the constellation Aquila where a dark streak in the Milky Way is also visible to the naked eye. The location of the Sun is indicated by the larger star in the interior of the diagram, slightly to the right of center. Thus, Herschel's star-gages indicated that our Solar System lay very near the center of the Milky Way system.

## Exploring Virtual Milky Ways

Herschel's star-gage method, and its limitations, can be explored in greater depth using the *Herschel's Star Gages* Easy Java Simulations model, which is available for free from the Open Source Physics collection on ComPADRE.[5] This computer program allows the user to observe stars fields in several virtual star systems. The view through the telescope, with a count of the number of stars in the visible field, is displayed in one window. A slider allows the user to rotate the telescope through 360 degrees in order to map out a cross section of the star system using Herschel's method.

While it is instructive to record star counts and compute the distance to the edge of the system by hand,[6] the program can also display a window that automatically plots the star-gages as the user moves the telescope around. The program can also display another window that shows a 3D view of the star system as seen from the outside. In this 3D view the program can show the cone of space that is visible to the telescope, and highlight the stars within that cone.

Figure 5 illustrates one of these virtual star systems (labeled Uniform 1 in the program). The left part of Figure 5 shows the 3D view of the star system, with the cone of view in yellow and stars in the telescope field highlighted in red. This star system is a cylindrical disk, with stars distributed uniformly throughout, and the telescope sweeps along the plane of the disk. The red points in the plot on the right of Figure 5 show the star-gage results. The star-gages clearly show the circular cross-section of the system, with the Sun (at the origin of the plot) off-center. Herschel's method works well in this case, because both of his fundamental assumptions hold.

But what happens if Herschel's assumptions are violated? The *Herschel's Star Gages* program allows the user to explore the consequences of violating assumption 2 above by limiting the viewing distance of the telescope. The green points in the star-gage plot of Figure 5 shows the result of allowing the telescope to only detect stars within 15 stellar units of the Sun.  In this case we see that the Sun appears to lie at the center of a circular system of stars with a radius of 15 stellar units.  In other words, having a limited sight distance makes us appear to be at the center of the galaxy.[7]

What about violations of assumption 1 above?  To let users explore this scenario, *Herschel's Star Gages* includes virtual star systems with non-uniform distributions of stars.  Figure 6 illustrates one such system (labeled Nonuniform 1 in the program).  The left part of Figure 6 shows the 3D view of the star system.  This system is just like the previous one except the stars are highly concentrated near the center of the disk. The plot on the right side of Figure 6 shows the star-gages: red points show the results with unlimited sight distance and green points show the results for a limiting distance of 10 stellar units.  These results indicate that the star-gages overestimate the distance to the galactic edge when the telescope is pointed toward a region of high star density, and underestimate the distance to the edge when the telescope is pointed toward a region of low star density.  The distance limited star-gage shows the Sun near the center of the system, but slightly displaced in the direction away from the region of highest star density.

## Evaluating Herschel's Star-Gages

The *Herschel's Star Gages* EJS model illustrates that Herschel's method can give accurate results for the shape of the galaxy, provided his two fundamental assumptions are valid.  But the simulation also shows that a limited sight distance can lead to results that make us appear to be located near the center of the Milky Way system even if we are not.  Likewise, deviations from uniform star density can lead to results that distort the true shape of the Milky Way, exaggerating distances toward high-density regions and underestimating distances toward low-density regions.

Looking back at Figure 2 we see that there are two possible interpretations for Herschel's results.  If his fundamental assumptions are valid, then Figure 2 shows an accurate cross-section of our galaxy.  However, Herschel's results can also be explained by assuming that his telescope could only see to a limited distance.  Perhaps the Milky Way extends much farther along the galactic plane than is indicated in Herschel's diagram.  If the results are due to a limited sight distance, then we might expect the Sun to be in the exact center of the diagram, rather than a bit off-center to the right.  But we have seen that this asymmetry can be explained if we assume that the star density is higher toward the left side of the diagram (in the direction of Aquila) than toward the right side (toward Canis Major).

Before the end of his career, Herschel recognized the potential problems with his star-gages.  In an 1817 paper he discussed star counts in the Milky Way using telescopes with different apertures.[8]  He found that larger aperture telescopes always revealed new stars that were not seen with smaller apertures, and in every case there appeared a whitish haze that likely indicated unresolved stars beyond the limits of his telescope.  He concluded that "By these observations it appears that the utmost stretch of the space-penetrating power of the 20 feet telescope could not fathom the Profundity of the milky way…" and even his 40-foot telescope could not see all the way to the edge.  Likewise, he recognized that the numerous star clusters he had found undermined the assumption of uniformity, so that "with regard to these gages, which on a supposition of an equality of scattering were looked upon as gages of distances, I have now to remark that, although a greater number of stars in the field of view is generally an indication of their greater distance from us, these gages, in fact, relate more immediately to the scattering of stars…"

We now recognize that our galaxy contains a layer of dust along the galactic plane.  This dust obscured Herschel's view of distant stars, even through his largest telescopes, thus violating assumption 2.  The Sun is not, in fact, close to the center of the Milky Way galaxy.  Furthermore, the cleft in the left side of Figure 2 is entirely an artifact of dust obscuring stars near the galactic plane.  Likewise, we now know that star densities are higher near the galactic center.  This explains why Herschel's diagram extends farther from the Sun to the left (toward Aquila, about 50 degrees away from the galactic center in Sagittarius) than to the right (toward Canis Major, about 130 degrees from the galactic center).

An examination of Herschel's star-gages provides insight into the important work of one of history's greatest astronomers, but it also illustrates an important aspect of the nature of science.  Scientists conduct their investigations of nature by making bold but reasonable assumptions that allow them to interpret the data they gather.  Without these assumptions it would be impossible for the investigation to proceed.  But the validity of the results hinge upon the validity of the assumptions.  The best scientists, like Herschel, do not hesitate to make use of reasonable assumptions, but they also work to test those assumptions, even long after their investigations are complete.

Readers interested in learning more about William Herschel are encouraged to consult the work of Herschel scholar Michael Hoskin.[9]  Instructors interested in using the *Herschel Star Gages* simulation in the classroom will find a classroom activity packaged with the simulation, or available as a separate download (see Ref. 5).

**Figures**

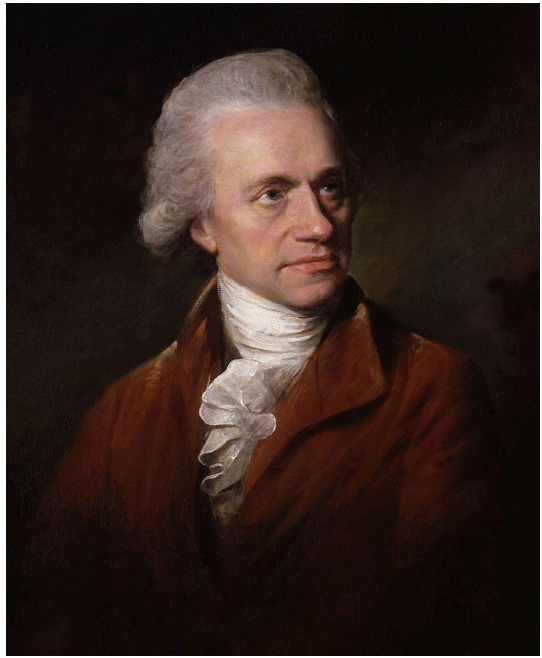

Figure 1. Portrait of William Herschel by L. Abbott (1785). National Portrait Gallery, London.

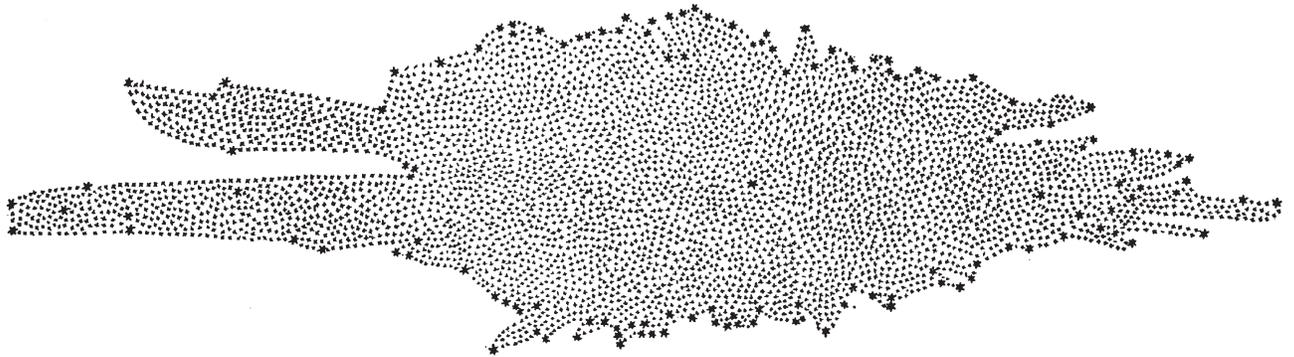

Figure 2. Herschel's cross-section of the Milky Way, as determined by his star-gages. From Ref. 4.

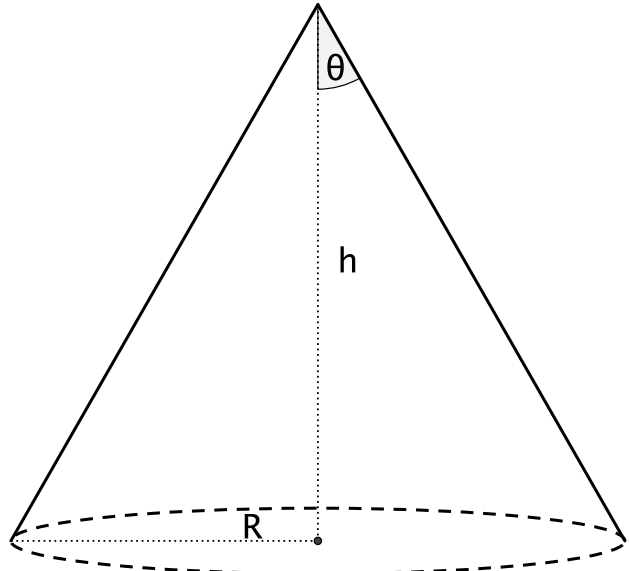

Figure 3. A cone with apex angle θ and height *h*.

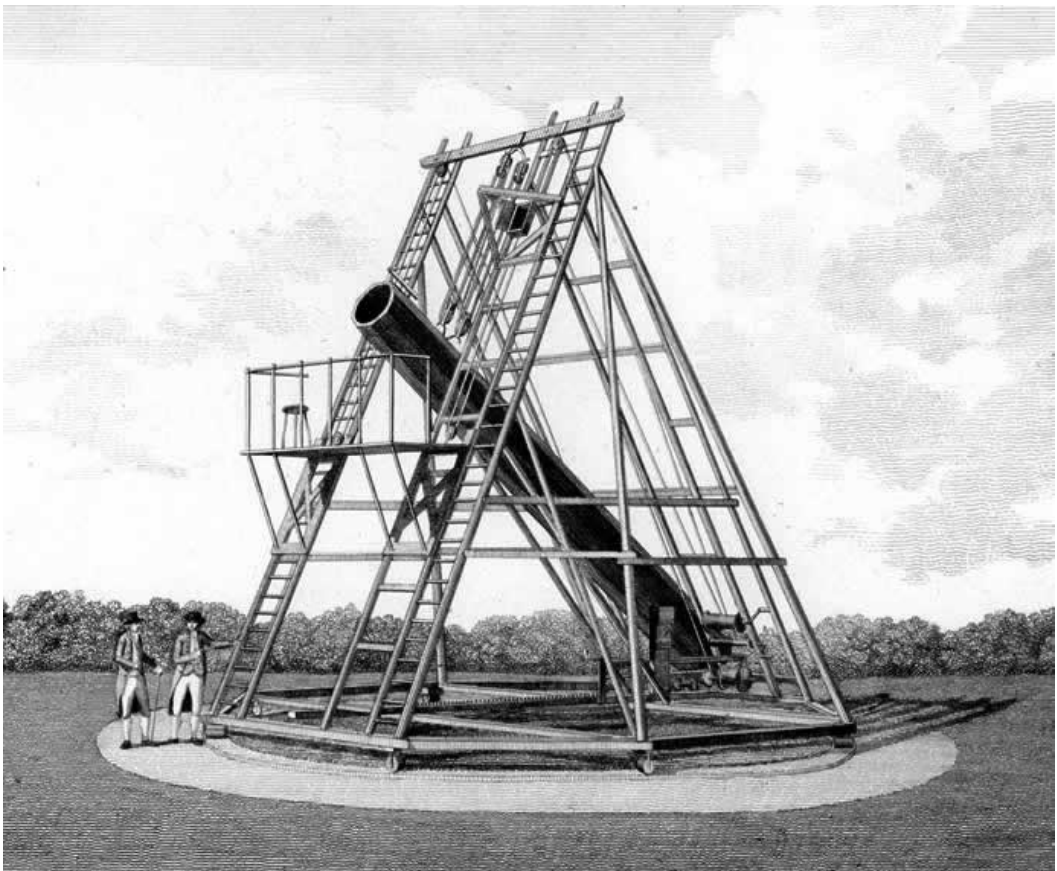

Figure 4. Herschel's "large" 20-foot telescope. Woodcut courtesy the Royal Astronomical Society.

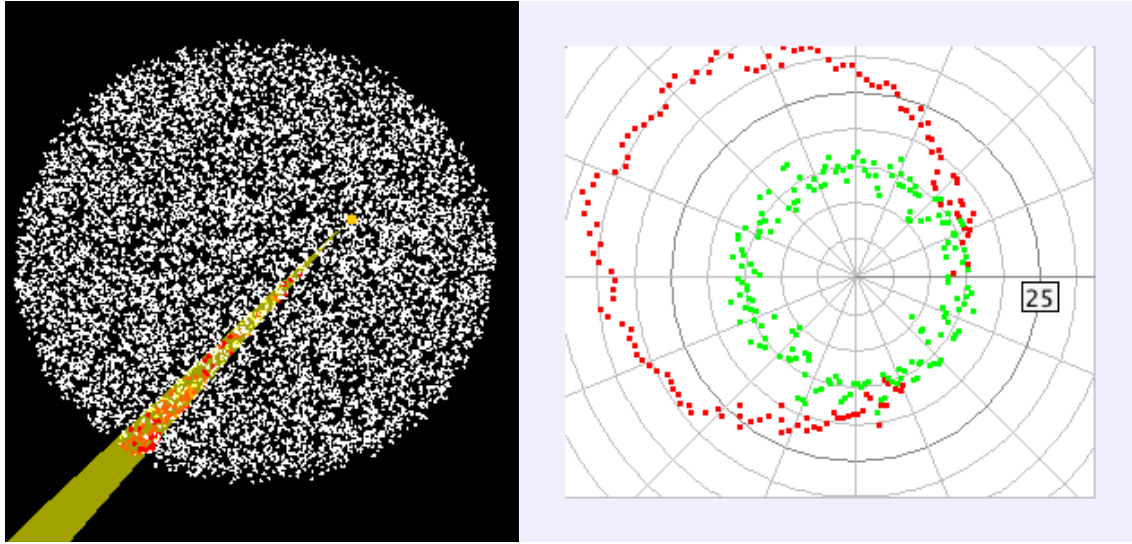

Figure 5. Uniform distribution of stars with sun off-center. The left image shows the system as seen from outside. The plot on the right shows the result of star-gages with unlimited distance (red) and with distance limited to 15 stellar units (green). Note that some of the green points lie on top of red points.

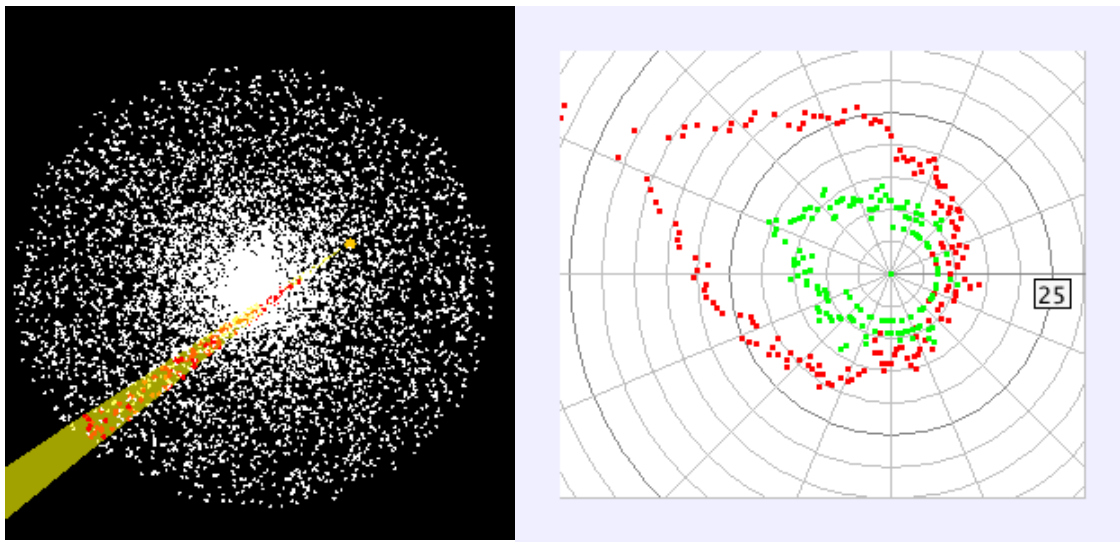

Figure 6. Non-uniform distribution of stars with sun off-center. The left image shows the system as seen from outside. The plot on the right shows the result of star-gages with unlimited distance (red) and with distance limited to 10 stellar units (green). Note that some of the green points lie on top of red points.